\documentclass[%
preprint,
 amsmath,amssymb,
]{revtex4-2}

\usepackage{graphicx}
\usepackage{dcolumn}
\usepackage{bm}
\usepackage[mathlines]{lineno}

\begin{document}


\title{Crossed Ponderomotive Lenses for Spherical Aberration Correction in Electron Optics}

\author{Yuuki Uesugi}
    \email{uesugi@tohoku.ac.jp}
    \affiliation{Institute of Multidisciplinary Research for Advanced Materials, Tohoku University, Katahira 2-1-1, Aoba-ku, Sendai 980-8577, Japan}
\author{Yuichi Kozawa}
    \affiliation{Institute of Multidisciplinary Research for Advanced Materials, Tohoku University, Katahira 2-1-1, Aoba-ku, Sendai 980-8577, Japan}

\date{April 9, 2025}

\begin{abstract}
This article evaluates the lens characteristics of a non-rotationally symmetric electron lens based on a ponderomotive potential (i.e., a ponderomotive lens) formed by intersecting one or more optical beams perpendicular to an electron beam.
Based on geometric optics, design formulas are derived for the focal length and general spherical aberration coefficients of specifically crossed ponderomotive lenses.
Numerical calculations demonstrate that a pair of these crossed ponderomotive lenses can effectively correct spherical aberration in the objective lens of an electron microscope.
Unlike rotationally symmetric ponderomotive lenses, which require the optical beam to be coaxially aligned with the electron beam, the crossed ponderomotive lens avoids the need to place optical mirrors and lenses directly on the beam axis.
Thus, it offers practical advantages in designing and building electron optical instruments and contributes to system miniaturization.
With lens properties similar to multipole lenses, the proposed crossed ponderomotive lens is expected to facilitate diverse developments in electron optical systems incorporating ponderomotive potentials.
\end{abstract}

\maketitle

\section{Introduction}
Electron lens systems constitute a core component of electron microscopes and have been studied for approximately 90 years to enhance nanoscale imaging and analysis.
Rotationally symmetric round lenses composed of electrodes or magnets are significantly constrained in shape compared to optical lenses.
Under paraxial conditions—where the electron velocity remains constant across the electron lens—the lens cannot provide negative lens power, and there are also strict constraints on the signs of geometrical and chromatic aberration coefficients \cite{Scherzer1936}.
Spherical aberration correction has long been the primary obstacle to achieving atomic resolution in electron microscopy. Although many correction methods have been proposed using rotationally symmetric designs \cite{Kawasaki2016,McMorran2017,Konecna2020,Schreiber2024}, the only practical solution so far has been to use combinations of non-rotationally symmetric multipole lenses.

In beam optics for relativistic charged particles, only multipole lenses are used for beam focusing, as required by the principle of strong focusing \cite{Snyder1952}, whereas rotationally symmetric round lenses are not used.
In electron microscopy, research on multipole lenses has advanced alongside spherical aberration correction techniques to compensate for the design limitations of round lenses \cite{Scherzer1947,Hawkes2009}.
However, accurately maintaining symmetry while precisely controlling the arrangement and excitation of multiple electrodes or magnets in multipole lenses posed significant technical challenges.
It was not until the late 1990s that researchers and engineers overcame these challenges, achieving spherical aberration correction and sub-angstrom resolution \cite{Haider1995,Krivanek1999,Batson2002,Nellist2004}.

About two decades after aberration correctors became practical, a renewed interest emerged in electron optics, particularly in the manipulation of non-relativistic electron beams.
When free electrons interact with optical fields exhibiting steep intensity gradients, such as optical standing waves, their momentum changes \cite{Kapitza1933,Herts1957,Schwarz1965,Bartell1965,Fedorov1967,Takeda1968,Bartell1968,Ezawa1969,Chan1979,Bucksbaum1988}.
The effective potential from this interaction, acting on electrons after time averaging over optical periods, is called the ponderomotive potential \cite{Fedorov1997}.
Recently, it has become widely recognized that this potential can serve as a new electron-optical element for manipulating electron beams \cite{Batelaan2001,Batelaan2002,Batelaan2007,Kozak2018,Muller2019}.
Applications to electron lens systems have proposed placing structured optical beams with rotationally symmetric intensity distributions coaxially with electron beams to achieve convex/concave lensing actions and spherical aberration correction \cite{GarciadeAbajo2021,Uesugi2021,Mihaila2025}.
Moreover, lensing action was experimentally demonstrated using a scanning electron microscope operated at an acceleration voltage of 30 kV and a near-infrared femtosecond laser with pulse energy of 30 \textmu J \cite{Mihaila2022}.

We have previously reported geometric optical analyses of rotationally symmetric electron lenses based on ponderomotive potentials (ponderomotive lenses) \cite{Uesugi2022}.
In this work, we extend the analysis to non-rotationally symmetric ponderomotive lenses exhibiting twofold and fourfold mirror symmetry.
Unlike rotationally symmetric designs that require coaxial alignment of optical and electron beams, our approach eliminates the need to place optical elements such as mirrors and focusing lenses directly on the electron beam axis, thereby minimizing beam disturbance and reducing system length.
This makes the method well-suited for integration into existing instruments and contributes to overall system miniaturization.
Following the geometric optics framework based on variational principles \cite{Hawkes2018, Uesugi2023}, we analyze the behavior of these non-rotationally symmetric ponderomotive lenses, treated as distributed media without refractive surfaces.
We propose several lens configurations, derive design formulas for their focal lengths and spherical aberration coefficients, and provide concrete parameters for correcting spherical aberration in electron microscope objective lenses.

\section{Cylindrical and Crossed Lenses}
We define the electron beam propagation direction as $z$, and introduce Cartesian coordinates $x$ and $y$ in the plane perpendicular to the $z$-axis, which originates at the beam axis.
The relationship between the ponderomotive potential $U$ and the optical intensity distribution $I$ is given by:
\begin{align}
    U(x,y,z) = \frac{2\pi r_e}{c k^2} I(x,y,z),
    \label{eq:U}
\end{align}
where, $r_e$, $c$, $k = 2\pi/\lambda$, and $\lambda$ denote the classical electron radius, speed of light, wavenumber, and wavelength, respectively.
Taking the gradient of this potential yields the force acting on the electron (the ponderomotive force), which is directed from regions of higher to lower optical intensity.

We first consider an intensity distribution symmetric about the $x$- or $y$-axis, corresponding to a cylindrical lens profile.
Then, by superposing two such beams orthogonally, we construct a cross-shaped lens with additional reflection symmetry along the $\pm45^\circ$ axes.
Unlike quadrupole lenses formed by electrostatic or magnetic fields, the crossed ponderomotive lens can provide lens power of the same sign in both the $xz$ and $yz$ planes.

\subsection{Cylindrical Lenses}
\begin{figure}[ht]
\centering
\includegraphics*[width=0.4\textwidth]{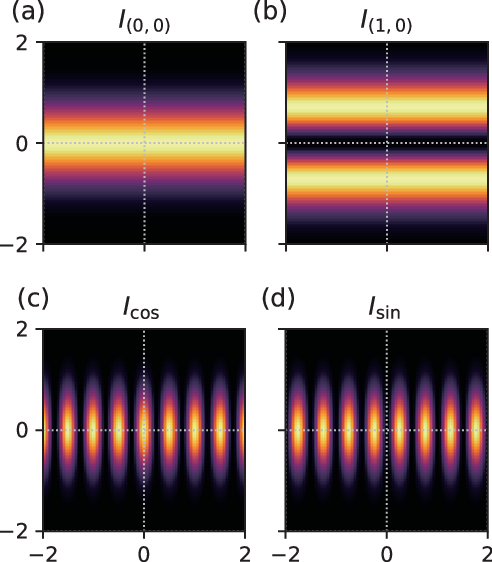}
\caption{
Intensity profiles in the $xy$ plane at $z=0$ for optical beams generating cylindrical lens effects.
(a) and (b) are based on HG beams, while (c) and (d) are based on standing waves. Calculated with $\lambda = w_0 = 1$.
}
\label{fig:beams}
\end{figure}

To realize a cylindrical lens for electron beams that exhibits lens power in only the $xz$ or $yz$ plane, we consider intensity profiles generated by Hermite-Gaussian (HG) beams and optical standing waves.
Let us first examine the HG beam propagating in the $x$ direction.
For electric field amplitude $E_0$, the HG beam is given by \cite{Yariv2007}
\begin{align}
    E_{(n,m)}(x,y,z)
    &= E_0\frac{w_0}{w(x)} H_n\left( \frac{\sqrt{2}y}{w(x)} \right) H_m\left( \frac{\sqrt{2}z}{w(x)} \right) \exp\left( -\frac{y^2 + z^2}{w^2(x)} \right) \nonumber\\
    &\times \exp\left( - ikx - i\frac{k(y^2 + z^2)}{2R(x)} + i(n+m+1)\zeta(x) \right),
    \label{eq:HG_E}
\end{align}
where
\begin{align}
    w(x) = w_0 \sqrt{1 + \left( \frac{x}{z_R} \right)^2}, \quad R(x) = x \left( 1 + \left( \frac{z_R}{x} \right)^2 \right), \quad \zeta(x) = \tan^{-1} \left( \frac{x}{z_R} \right),
\end{align}
and $z_R = \pi w_0/\lambda$ is the Rayleigh range.
We assume the electron beam enters at the waist center of the HG beam ($x=0$), where $w(x) \approx w_0$ holds.
The intensity distributions for the fundamental ($n=m=0$) and first-order ($n=1, m=0$) modes are
\begin{align}
    I_{(0,0)}(x,y,z) &= I_0 \exp\left( -\frac{2(y^2 + z^2)}{w_0^2} \right), \label{eq:HG00_I} \\
    I_{(1,0)}(x,y,z) &= \frac{8I_0 y^2}{w_0^2} \exp\left( -\frac{2(y^2 + z^2)}{w_0^2} \right), \label{eq:HG10_I}
\end{align}
where $I_0 = |E_0|^2/2\eta$ and $\eta = 377\;\mathrm{\Omega}$.
The relationship between $I_0$ and the input optical power $P = \int I(x,y,z)\,dydz$ depends on the beam profile:
\begin{align}
    I_0 =
        \begin{cases}
        \dfrac{2 P}{\pi w_0^2} & \mathrm{for} \; I_{(0,0)}, \\
        \dfrac{P}{\pi w_0^2} & \mathrm{for} \; I_{(1,0)}.
        \end{cases}
    \label{eq:I0}
\end{align}
Figures \ref{fig:beams}a and b show the intensity profiles in the $xy$ plane at $z=0$ for Eqs. (\ref{eq:HG00_I}) and (\ref{eq:HG10_I}).
The beam of Eq. (\ref{eq:HG00_I}), with peak intensity on the beam axis, produces a diverging effect in the $yz$ plane.
If an electron beam with a spot size smaller than the beam waist is injected into the dark-centered profile of Eq. (\ref{eq:HG10_I}), it experiences a focusing effect.
Higher-order HG beams ($n\geq2$) show intensity profiles with bright or dark lines along the $x$ axis depending on the parity of $n$, but the resulting lens characteristics are similar to the $n=0$ and $n=1$ cases and are omitted here.

Next, we describe the cylindrical lens profile formed by optical standing waves, generated by two counterpropagating fundamental Gaussian beams along the $x$ axis:
\begin{align}
    E_{\mathrm{sw}}(x,y,z)
    \approx 2 E_0 \exp\left( -\frac{2(y^2 + z^2)}{w_0^2} \right) \cos(kx).
\end{align}
Its intensity distribution is
\begin{align}
    I_{\mathrm{cos}}(x,y,z)
    = 4I_0 \exp\left( -\frac{2(y^2 + z^2)}{w_0^2} \right) \cos^2 (kx).
\end{align}
This produces a bright line at $x=0$ and acts as a diverging cylindrical lens in the $xz$ plane (Fig. \ref{fig:beams}c).
Replacing $\cos$ with $\sin$ gives
\begin{align}
    I_{\mathrm{sin}}(x,y,z) = 4I_0 \exp\left( -\frac{2(y^2 + z^2)}{w_0^2} \right) \sin^2 (kx),
\end{align}
which forms a dark line at $x=0$ and provides focusing action in the $xz$ plane (Fig. \ref{fig:beams}d).

\subsection{Crossed Lenses}
\begin{figure}[ht]
\centering
\includegraphics*[width=0.4\textwidth]{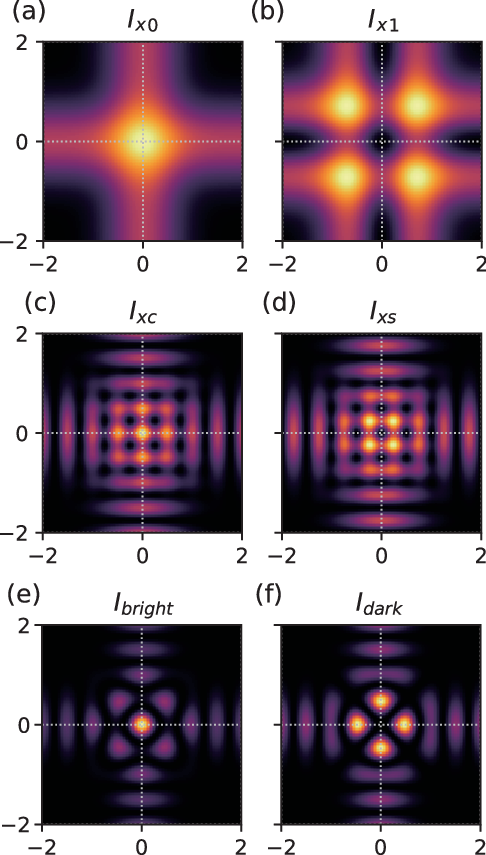}
\caption{
Intensity profiles in the $xy$ plane at $z=0$ for optical fields generating crossed lens effects.
(a) and (b) are constructed by incoherently superposing HG beams, (c) and (d) by incoherently superposing standing waves, and (e) and (f) by coherently superposing standing waves to form optical lattices.
Calculated with $\lambda = w_0 = 1$.
}
\label{fig:cross}
\end{figure}

We define a crossed lens as the superposition of two identical cylindrical lenses arranged orthogonally in the $xy$ plane.
Incoherent superposition of beams described by Eqs. (\ref{eq:HG00_I}) and (\ref{eq:HG10_I}) yields
\begin{align}
    I_{\mathrm{x0}}(x,y,z) &= I_{(0,0)}(x,y,z) + I_{(0,0)}(y,x,z), \\
    I_{\mathrm{x1}}(x,y,z) &= I_{(1,0)}(x,y,z) + I_{(1,0)}(y,x,z),
\end{align}
which form bright and dark spots at the center of the $xy$ plane, respectively, and act as diverging and focusing crossed lenses (Figs. \ref{fig:cross}a and b).
In contrast, coherent superposition of HG beams produces interference fringes oriented at $\pm45^\circ$, resulting only in diagonally tilted cylindrical lenses---unsuitable for crossed lens formation.

Similarly, incoherent superposition of orthogonal standing waves gives
\begin{align}
    I_{\mathrm{xc}}(x,y,z) &= I_{\mathrm{cos}}(x,y,z) + I_{\mathrm{cos}}(y,x,z), \\
    I_{\mathrm{xs}}(x,y,z) &= I_{\mathrm{sin}}(x,y,z) + I_{\mathrm{sin}}(y,x,z),
\end{align}
which also form central bright and dark spots and act as diverging and focusing crossed lenses (Figs. \ref{fig:cross}c and d).
Unlike the case of HG beams, coherent superposition of standing waves yields two types of optical lattices:
\begin{align}
    I_{\mathrm{bright}}(x,y,z)
    &= \frac{\left| E_{\mathrm{sw}}(x,y,z) + E_{\mathrm{sw}}(y,x,z) \right|^2}{2 \eta}, \nonumber\\
    &= 4 I_0 \left[ \exp\left( -\frac{2(y^2 + z^2)}{w_0^2} \right) \cos(kx) + \exp\left( -\frac{2(x^2 + z^2)}{w_0^2} \right) \cos(ky) \right]^2, \\
    I_{\mathrm{dark}}(x,y,z)
    &= \frac{\left| E_{\mathrm{sw}}(x,y,z) - E_{\mathrm{sw}}(y,x,z) \right|^2}{2 \eta}, \nonumber\\
    &= 4 I_0 \left[ \exp\left( -\frac{2(y^2 + z^2)}{w_0^2} \right) \cos(kx) - \exp\left( -\frac{2(x^2 + z^2)}{w_0^2} \right) \cos(ky) \right]^2.
\end{align}
These intensity distributions in the $xy$ plane at $z=0$ are shown in Figs. \ref{fig:cross}e and f.
$I_{\mathrm{bright}}$ represents a bright optical lattice on the $z$ axis, while $I_{\mathrm{dark}}$ represents a dark one.

\section{Defocus and Spherical Aberration of Crossed Lenses}
We consider the propagation of an electron beam through an anamorphic system composed of cylindrical lenses acting in the $xz$ or $yz$ plane.
The action integral for the electron beam, under the electron optical approximation, is given by
\begin{align}
    S = \int  \mathbf{p} \cdot d \mathbf{s} = \int  p(x,y,z)  \sqrt{1+x'^2 + y'^2}  dz,
\end{align}
where $d\mathbf{s}$ is the line element along the trajectory; the prime ($'$) indicates $d/dz$.
The canonical momentum $p$ is a function that includes the shape of the ponderomotive potential.
Since we assume an anamorphic system, the function $p$ has two invariants $x^2$ and $y^2$.
Expanding $p$ up to fourth order around $x=y=0$, we obtain
\begin{align}
    p = p_{0} + p_{10}  x^2 + p_{01}  y^2 + p_{20}  x^4 + p_{11}  x^2 y^2 + p_{02}  y^4,
\end{align}
where the coefficients $p_{ij}$ are generally functions of $z$.
The integrand of the action integral defines the variational function,
\begin{align}
    F(x,y,x',y',z)=p(x,y,z) \sqrt{1+x'^2 + y'^2},
\end{align}
where the coefficient of $p$ depends on the invariant $r'^2 = x'^2 + y'^2$.
Expanding $F$ to fourth order yields
\begin{align}
    F
    &= p_0 + p_{10} x^2 + p_{01} y^2 + \frac{p_0}{2} r'^2 \nonumber\\
    &+ p_{20} x^4 + p_{11} x^2 y^2  + p_{02} y^4 + \frac{p_{10}}{2} x^2 r'^2 + \frac{p_{01}}{2} y^2 r'^2 - \frac{p_0}{8} r'^4.
\end{align}

If the lens system consists solely of crossed lenses (crossed lens system), the system exhibits symmetry under the exchange of the $xz$ and $yz$ planes.
The fourth-order expansion of the momentum then becomes
\begin{align}
    p = p_{0} + p_1(x^2 + y^2) + p_2(x^4 + y^4) + p_{11} x^2 y^2,
\end{align}
where the new expansion coefficients relate to those in the previous expression via
\begin{align}
    p_1 &= p_{10} = p_{01},\\
    p_2 &= p_{20} = p_{02}.
\end{align}
Using these relations, the fourth-order expansion of the variational function in the crossed lens system becomes
\begin{align}
    F
    &= p_0 + p_1 (x^2 + y^2) + \frac{p_0}{2} r'^2 \nonumber\\
    &+ p_2 (x^4 + y^4) + p_{11} x^2 y^2 + \frac{p_1}{2} (x^2 + y^2)r'^2 - \frac{p_0}{8} r'^4.
\end{align}
Compared to the fourth-order expansion for a rotationally symmetric system, the cross term $x^2 y^2$ is the key difference.

\subsection{Paraxial Rays}
Paraxial rays in an anamorphic system can be derived from the second-order terms, forming two independent trajectories in the $xz$ and $yz$ planes \cite{Yuan2009-1}.
Applying the thin-lens approximation as in \cite{Uesugi2022, Uesugi2023}, the lens powers in the $xz$ and $yz$ planes are given by
\begin{align}
    \frac{1}{f_x} = -2\int_{\mathrm{lens}} \frac{p_{10}}{p_0} dz,\quad \frac{1}{f_y} = -2\int_{\mathrm{lens}} \frac{p_{01}}{p_0} dz,
\end{align}
where $f_x$ and $f_y$ are the focal lengths in the $xz$ and $yz$ planes, respectively.
In an anamorphic system, the chief ray and marginal ray are independently defined in the $xz$ and $yz$ planes.
Let the object coordinates be $(x_o, y_o)$, the aperture coordinates be $(x_a, y_a)$, and the corresponding rays be $s_x, s_y$ and $t_x, t_y$.
Any ray in the $xz$ or $yz$ plane is a linear combination:
\begin{align}
    x = x_o s_x + x_a t_x, \quad
    x' = x_o s'_x + x_a t'_x, \quad
    y = y_o s_y + y_a t_y, \quad
    y' = y_o s'_y + y_a t'_y,
\end{align}

For the crossed lens system, there is only a single second-order term $r^2 = x^2 + y^2$, so paraxial rays behave identically to those in a rotationally symmetric system.
The focal length $f$ of the crossed lens is common in both $xz$ and $yz$ planes and is given by
\begin{align}
    \frac{1}{f} &= -2\int_{\mathrm{lens}} \frac{p_1}{p_0} dz.
    \label{eq:lens_power}
\end{align}
The chief and marginal rays are defined by a single pair $s$ and $t$, and any ray between the object and image planes is given by
\begin{align}
    x = x_o s + x_a t, \quad
    x' = x_o s' + x_a t', \quad
    y = y_o s + y_a t, \quad
    y' = y_o s' + y_a t'.
\end{align}

\subsection{Primary Spherical Aberration}
The wavefront aberration is the optical path difference between the reference wave and the actual wavefront.
The primary wavefront aberration is derived from the fourth-order expansion term $F_4$ of the variational function as
\begin{align}
W(x_o,y_o,x_a,y_a) = D \int_{z_o}^{z_i} F_4(x_o,y_o,x_a,y_a;z) dz,
\label{eq:wave_abb}
\end{align}
where $D$ is a factor converting the action integral to optical path length, determined by ray propagation in drift space; for a ponderomotive lens, $D = -1/p_0$.
The wavefront aberration depends on object coordinates $(x_o, y_o)$ and aperture coordinates $(x_a, y_a)$, so $F_4$ must also be transformed into a function of these variables.
Discarding offset terms that do not contribute to aberrations, the transformed $F_4$ yields 16 terms in the case of the anamorphic system \cite{Yuan2009-1}.

When the object lies on the optical axis ($x_o = y_o = 0$), most aberration terms vanish.
The remaining fourth-order terms constitute what is commonly referred to as primary general spherical aberration:
\begin{align}
    F_{\mathrm{4s}}(x_a,y_a) = A_1x_a^4 + A_2y_a^4 + A_3x_a^2 y_a^2,
\end{align}
where coefficients are given by
\begin{align}
    A_1 &= p_{20}t_x^4 + \frac{p_{10}}{2}t_x^2 t_x'^2 - \frac{p_0}{8} t_x'^4, \\
    A_2 &= p_{02}t_y^4 + \frac{p_{01}}{2}t_y^2 t_y'^2 - \frac{p_0}{8} t_y'^4, \\
    A_3 &= p_{11}t_x^2 t_y^2 + \frac{1}{2}\left(p_{10}t_x^2 t_y'^2 + p_{01}t_y^2 t_x'^2\right) - \frac{p_0}{4} t_x'^2 t_y'^2.
\end{align}
In the crossed lens system, symmetry between $x$ and $y$ implies $A_1 = A_2 = A_0$, where
\begin{align}
    A_0 = p_2 t^4 + \frac{p_1}{2} t^2 t'^2 - \frac{p_0}{8} t'^4.
\end{align}
$A_3$ is expressed using $A_0$ as
\begin{align}
    A_3
    &= p_{11} t^4 + p_1 t^2 t'^2 - \frac{p_0}{4} t'^4 \nonumber\\
    &= 2 A_0 + A_4,
\end{align}
where
\begin{align}
    A_4 = (p_{11} - 2 p_2)t^4.
\end{align}
Therefore, the general spherical aberration in the crossed lens system is given by
\begin{align}
    F_{\mathrm{4s}}(x_a,y_a) = A_0 r_a^4 + A_4 x_a^2 y_a^2,
\label{eq:4s}
\end{align}
with $r_a^2 = x_a^2 + y_a^2$.
The coefficient $p_{11} - 2p_2$ produces a cross term that distinguishes the crossed lens system from a rotationally symmetric one.

\subsection{General Spherical Aberration Coefficients}
In electron optics, aberrations are usually discussed in terms of ray aberrations at the object plane, defined using the wavefront aberration as
\begin{align}
    \Delta\mathbf{r} = 
    \left( \begin{array}{cc}
        \Delta x \\
        \Delta y \\
    \end{array} \right)
    = \frac{z_i - z_a}{M}\mathbf{\nabla}_a W(x_o,y_o,x_a,y_a),
    \label{eq:ray_abb}
\end{align}
where $\mathbf{\nabla}_a$ is the gradient with respect to aperture coordinates, and $M$ is the transverse magnification of the lens system.
Note that in general anamorphic systems, a single common $M$ for both $x$ and $y$ components doses not exist. Substituting Eq. (\ref{eq:4s}) into Eq. (\ref{eq:ray_abb}) yields
\begin{align}
    \Delta x = (\bar{A}_0 r_a^2 + \bar{A}_4 y_a^2)x_a, \\
    \Delta y = (\bar{A}_0 r_a^2 + \bar{A}_4 x_a^2)y_a,
\end{align}
where
\begin{align}
    \bar{A}_0 &= -\frac{4(z_i - z_a)}{p_0 M}\int_{z_o}^{z_i} A_0 dz,\\
    \bar{A}_4 &= -\frac{2(z_i - z_a)}{p_0 M}\int_{z_o}^{z_i} A_4 dz.
\end{align}
Using polar coordinates, $x_a = r_a\cos\theta$ and $y_a = r_a\sin\theta$, the ray aberration becomes
\begin{align}
    \Delta \mathbf{r}
    &= \bar{A}_0 r_a^3
    \left( \begin{array}{cc}
        \cos\theta \\
        \sin\theta \\
    \end{array} \right)
    + \frac{\bar{A}_4 r_a^3}{2} \sin 2\theta
    \left( \begin{array}{cc}
        \sin\theta\ \\
        \cos\theta\ \\
    \end{array} \right), \nonumber\\
    & = \Delta \mathbf{r}_0 + \Delta \mathbf{r}_4.
    \label{eq:r0_r4}
\end{align}
The first term $\Delta \mathbf{r}_0$ corresponds to spherical aberration in a rotationally symmetric system.
The second term $\Delta \mathbf{r}_4$ creates a fourfold symmetric aberration.
We refer to these as the circular and four-lobed components, respectively.

For practical purposes, we redefine the ray aberration using the object and image planes.
To eliminate aperture dependence, the marginal ray $t$ is expressed in terms of the fundamental solution $h(z)$ satisfying $(r(z_o), r'(z_o)) = (0, 1)$, with $z_o$ denoting the object plane position.
Since the aperture plane can be arbitrarily placed between the object and image planes, we locate it at the lens position.
Assuming shallow ray angles, we obtain
\begin{align}
t = \frac{h}{a}, \quad \frac{r_a}{a} = \alpha, \quad \frac{z_i - z_a}{a} = \frac{b}{a} = M,
\end{align}
where $\alpha$ is the object-side aperture angle, $a$ is the distance from object to lens, and $b$ is from lens to image.
Substituting these into Eq. (\ref{eq:r0_r4}), we obtain
\begin{align}
    \Delta \mathbf{r}_0
    &= C_{\mathrm{s0}} \alpha^3
    \left( \begin{array}{cc}
        \cos\theta \\
        \sin\theta \\
    \end{array} \right),
    \\
    \Delta \mathbf{r}_4
    &= C_{\mathrm{s4}} \alpha^3  \sin 2\theta
    \left( \begin{array}{cc}
        \sin\theta\ \\
        \cos\theta\ \\
    \end{array} \right),
\end{align}
with coefficients
\begin{align}
    C_{\mathrm{s0}} &= -\frac{4}{p_0}\int_{z_o}^{z_i} \left( p_2 h^4 + \frac{p_1}{2}h^2h'^2 - \frac{p_0}{8} h'^4 \right) dz, \label{eq:Cs0}\\
    C_{\mathrm{s4}} &= -\frac{1}{p_0}\int_{z_o}^{z_i} (p_{11} - 2 p_2)h^4 dz. \label{eq:Cs4}
\end{align}
These represent the coefficients of the circular and four-lobed components of ray aberration at the object plane.
Although $h'$ can sometimes be eliminated analytically once $p$ is specified, it is often difficult to do so in general practice.
Therefore, assuming thin lenses and small beam divergence, we apply a rough approximation by neglecting integrals involving $h'$; this approximation was used for rotationally symmetric ponderomotive lenses in \cite{Uesugi2022, Uesugi2023}.
The integration range is replaced with the lens thickness $l$ and the ray height is assumed constant at $h=a$ over this range.
Setting $z=0$ at the lens center, Eqs. (\ref{eq:Cs0}) and (\ref{eq:Cs4}) reduce to
\begin{align}
    C_{s0} &\approx -\frac{4 a^4}{p_0}\int_{-l/2}^{l/2} p_2 \,dz, \label{eq:Cs0_red} \\
    C_{s4} &\approx -\frac{a^4}{p_0}\int_{-l/2}^{l/2} (p_{11} - 2 p_2) \,dz. \label{eq:Cs4_red}
\end{align}
These are the simplified expressions for the general spherical aberration coefficients in the crossed lens system.

\section{Lens Properties and Spherical Aberration Correction}
Let the electron kinetic energy be $T$.
We define the modified interaction potential as
\begin{align}
    \hat{U} = \left( T - U \right)\left( 1 + \frac{T-U}{2mc^2} \right),
\end{align}
which gives the canonical momentum of the electron as
\begin{align}
    p = \sqrt{2m\hat{U}}.
\end{align}
Using these expressions along with Eq.~(\ref{eq:U}) and the six intensity distributions $I_{\mathrm{x0}}$, $I_{\mathrm{x1}}$, $I_{\mathrm{xc}}$, $I_{\mathrm{xs}}$, $I_{\mathrm{bright}}$, and $I_{\mathrm{dark}}$ derived in the previous section, the corresponding expansion coefficients of $p$ are summarized in Table \ref{table:p}.
The derivation uses the conditions $U_0 \ll T, mc^2$ and $1 \ll k^2 w_0^2$, which correspond to the assumptions that the electron beam and light beams are both paraxial.
In Table \ref{table:p}, $\gamma$ is the Lorentz factor, $p_0$ is given by $p_0 = \sqrt{m T (1 + \gamma)}$ for all cases, and $U_0$ is defined by replacing $I(x,y,z)$ in Eq.~(\ref{eq:U}) with $I_0$.
\begin{table}[ht]
    \centering
    \begin{tabular}{c|c c c c}
    \hline
    & $p_1$ & $p_2$ & $p_{11}$ \\
    \hline
    \hline
    $I_\mathrm{x0}$ &
        $\dfrac{2 U_0 \exp\left(-\frac{2 z^2}{w_0^2}\right)}{T w_0^2 (1 + \gamma)} p_0$ &
        $-\dfrac{1}{w_0^2} p_1$ &
        $-\dfrac{1}{\gamma^2} \dfrac{p_1^2}{p_0}$ \\
    $I_\mathrm{x1}$ &
        $-\dfrac{8 U_0 \exp\left(-\frac{2 z^2}{w_0^2}\right)}{T w_0^2 (1 + \gamma)} p_0$ &
        $-\dfrac{2}{w_0^2} p_1$ &
        $-\dfrac{1}{\gamma^2} \dfrac{p_1^2}{p_0}$ \\
    $I_\mathrm{xc}$ &
        $\dfrac{4 U_0 k^2 \exp\left(-\frac{2 z^2}{w_0^2}\right)}{T (1 + \gamma)} p_0$ &
        $-\dfrac{k^2}{3} p_1$ &
        $-\dfrac{1}{\gamma^2} \dfrac{p_1^2}{p_0} -\dfrac{4}{w_0^2} p_1$ \\
    $I_\mathrm{xs}$ &
        $-\dfrac{4 U_0 k^2 \exp\left(-\frac{2 z^2}{w_0^2}\right)}{T (1 + \gamma)} p_0$ &
        $-\dfrac{k^2}{3} p_1$ &
        $-\dfrac{1}{\gamma^2} \dfrac{p_1^2}{p_0} -\dfrac{4}{w_0^2} p_1$ \\
    $I_\mathrm{bright}$ &
        $\dfrac{8 U_0 k^2 \exp\left(-\frac{2 z^2}{w_0^2}\right)}{T (1 + \gamma)} p_0$ &
        $-\dfrac{5 k^2}{24} p_1$ &
        $\dfrac{6}{5} p_2$ \\
    $I_\mathrm{dark}$ &
        $0$ &
        $-\dfrac{U_0 k^4 \exp\left(-\frac{2 z^2}{w_0^2}\right)}{T (1 + \gamma)} p_0$ &
        $-2 p_2$ \\
    \hline
    \end{tabular}
    \caption{Expansion coefficients up to fourth order of the canonical momentum, $p$, in each crossed lens  derived in the previous section.}
    \label{table:p}
\end{table}

\subsection{Design formulas for Crossed Lens Properties}
By substituting the expansion coefficients from Table \ref{table:p} into Eq. (\ref{eq:lens_power}) and Eqs.(\ref{eq:Cs0_red}), (\ref{eq:Cs4_red}), design formulas for the lens power and general spherical aberration coefficients can be derived.
These results are summarized in Table \ref{table:property}, with the sign of the circular component $C_{\mathrm{s0}}$ indicated in parentheses for clarity.
The table shows that crossed lenses constructed by incoherent superposition of HG beams ($I_{\mathrm{x0}}, I_{\mathrm{x1}}$) and those made from incoherent superposition of optical standing waves ($I_{\mathrm{xc}}, I_{\mathrm{xs}}$) exhibit nearly identical properties.
Additionally, lenses with intensity maxima along the $z$-axis ($I_{\mathrm{x0}}, I_{\mathrm{xc}}$) act as concave lenses with positive circular components, while those with central dark regions ($I_{\mathrm{x1}}, I_{\mathrm{xs}}$) behave as convex lenses with negative circular components.
In contrast, crossed lenses formed from coherent superposition of standing waves exhibit distinct characteristics.
The bright lattice ($I_{\mathrm{bright}}$), with maxima along the $z$-axis, functions as a concave lens with a $C_{\mathrm{s4}} / C_{\mathrm{s0}}$ ratio of $-1/5$.
The dark lattice ($I_{\mathrm{dark}}$), forming a central minimum, lacks a defocusing term and does not act as a lens.
Its $C_{\mathrm{s4}} / C_{\mathrm{s0}}$ ratio is $-1$.
\begin{table}[ht]
    \centering
    \begin{tabular}{c|c c c c}
    \hline
    & $1/f$ & $C_{\mathrm{s0}}$ & $C_{\mathrm{s4}}$ \\
    \hline
    \hline
    $I_\mathrm{x0}$ &
        $-\dfrac{2\sqrt{2\pi} U_0}{T w_0 (1 + \gamma)}$ &
        $-\dfrac{2 a^4}{w_0^2 f} \;(+)$ &
        $-\dfrac{1}{2} C_{\mathrm{s0}}$ \\
    $I_\mathrm{x1}$ &
        $\dfrac{8\sqrt{2\pi} U_0}{T w_0 (1 + \gamma)}$ &
        $-\dfrac{4 a^4}{w_0^2 f} \;(-)$ &
        $-\dfrac{1}{2} C_{\mathrm{s0}}$ \\
    $I_\mathrm{xc}$ &
        $-\dfrac{4\sqrt{2\pi} U_0 k^2 w_0}{T (1 + \gamma)}$ &
        $-\dfrac{2 a^4 k^2}{3 f} \;(+)$ &
        $-\dfrac{1}{2} C_{\mathrm{s0}}$ \\
    $I_\mathrm{xs}$ &
        $\dfrac{4\sqrt{2\pi} U_0 k^2 w_0}{T (1 + \gamma)}$ &
        $-\dfrac{2 a^4 k^2}{3 f} \;(-)$ &
        $-\dfrac{1}{2} C_{\mathrm{s0}}$ \\
    $I_\mathrm{bright}$ &
        $-\dfrac{8\sqrt{2\pi} U_0 k^2 w_0}{T (1 + \gamma)}$ &
        $-\dfrac{5 a^4 k^2}{12 f} \;(+)$ &
        $-\dfrac{1}{5} C_{\mathrm{s0}}$ \\
    $I_\mathrm{dark}$ &
        $0$ &
        $\dfrac{2\sqrt{2\pi} U_0 a^4 k^4 w_0}{T (1 + \gamma)} \;(+)$ &
        $-C_{\mathrm{s0}}$ \\
    \hline
    \end{tabular}
    \caption{Lens power and general spherical aberration coefficients for each crossed lens derived in the previous section. For clarity, the sign of each $C_{\mathrm{s0}}$ entry is indicated in parentheses.}
    \label{table:property}
\end{table}

\begin{figure}[ht]
\centering
\includegraphics*[width=0.9\textwidth]{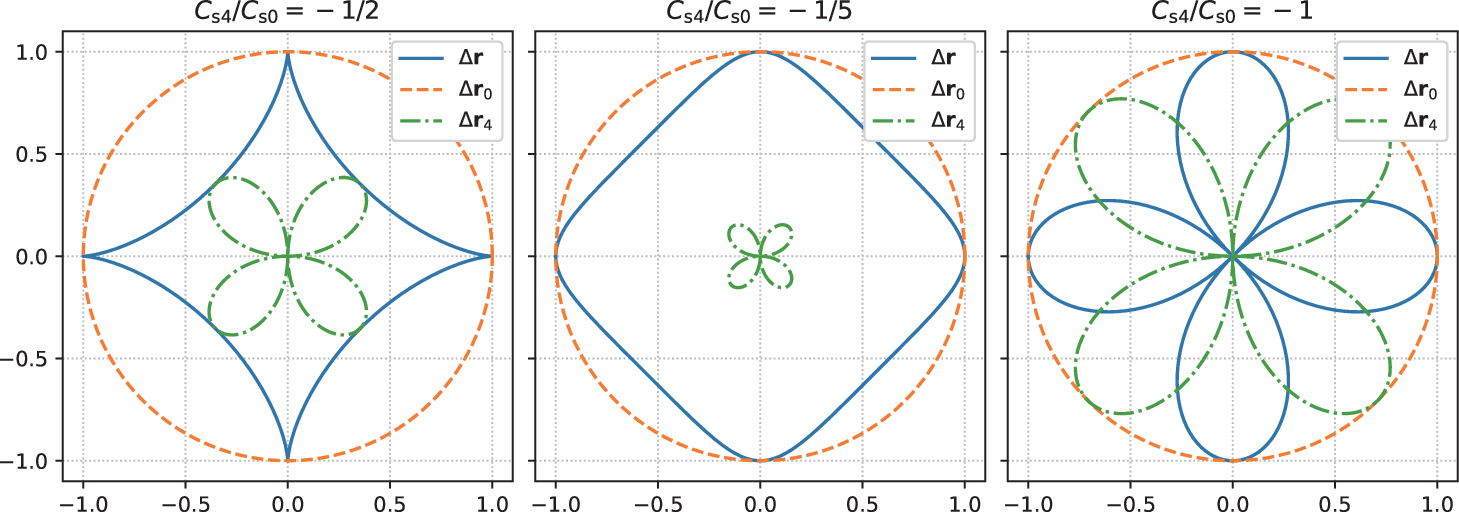}
\caption{Plots of general spherical aberration under the condition $C_{\mathrm{s0}}\alpha^3 = 1$, where $\Delta \mathbf{r} = \Delta \mathbf{r}_0 + \Delta \mathbf{r}_4$.}
\label{fig:diagram}
\end{figure}

Crossed lenses with the same value of $C_{\mathrm{s4}}/C_{\mathrm{s0}}$ exhibit wavefronts of general spherical aberration that are geometrically similar, aside from the overall sign.
Figure \ref{fig:diagram} shows plots of the ray aberration $\Delta \mathbf{r}$ and its circular and four-lobed components $\Delta \mathbf{r}_0$ and $\Delta \mathbf{r}_4$, respectively, for $C_{\mathrm{s4}}/C_{\mathrm{s0}} = -1/2$, $-1/5$, and $-1$.
These were obtained using Eqs. (\ref{eq:Cs0_red}) and (\ref{eq:Cs4_red}) under the condition $C_{\mathrm{s0}}\alpha^3 = 1$.
The first two cases yield diamond-shaped aberration patterns resulting from the sum of circular and four-lobed components.  
In contrast, when $C_{\mathrm{s4}}/C_{\mathrm{s0}} = -1$, the total aberration is equivalent to a four-lobed component rotated by 45 degrees in the $xy$ plane.  
This corresponds to the fact that the sum of a four-lobed component based on the $xy$ coordinates and one based on a coordinate rotated by 45 degrees yields a circular component.

\subsection{Example of Spherical Aberration Correction}
\begin{figure}[ht]
\centering
\includegraphics*[width=0.33\textwidth]{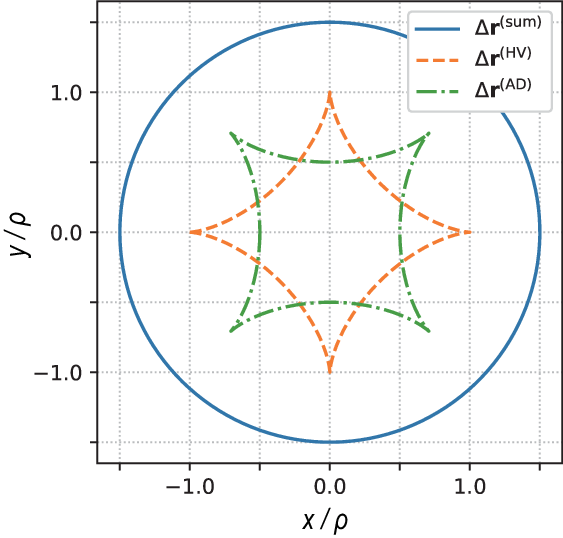}
\caption{General spherical aberration from a pair of crossed lenses, lens HV and lens AD.
The resulting aberration is rotationally symmetric and negative.}
\label{fig:correction}
\end{figure}

To correct the positive spherical aberration introduced by the objective lens of an electron microscope, it suffices to generate general spherical aberration with only a negative circular component.  
One way to achieve this is by using a pair of crossed lenses constructed from $I_{\mathrm{x1}}$ or $I_{\mathrm{xs}}$.
Let us denote the two crossed lenses as lens HV and lens AD, where HV is based on the standard $xy$ coordinates, and AD is based on a coordinate system rotated by 45 degrees.
Assuming $C_{\mathrm{s0}}\alpha^3 = -\rho$, Fig.~\ref{fig:correction} shows the ray aberrations of the lens HV, $\Delta\mathbf{r}^{\mathrm{(HV)}}$, the lens AD, $\Delta\mathbf{r}^{\mathrm{(AD)}}$, and their sum, $\Delta\mathbf{r}^{\mathrm{(sum)}} = \Delta\mathbf{r}^{\mathrm{(HV)}} + \Delta\mathbf{r}^{\mathrm{(AD)}}$.
The four-lobed components from the two lenses combine into a circular component, resulting in a rotationally symmetric spherical aberration with magnitude $-1.5\rho$.

\begin{figure}[ht]
\centering
\includegraphics*[width=0.5\textwidth]{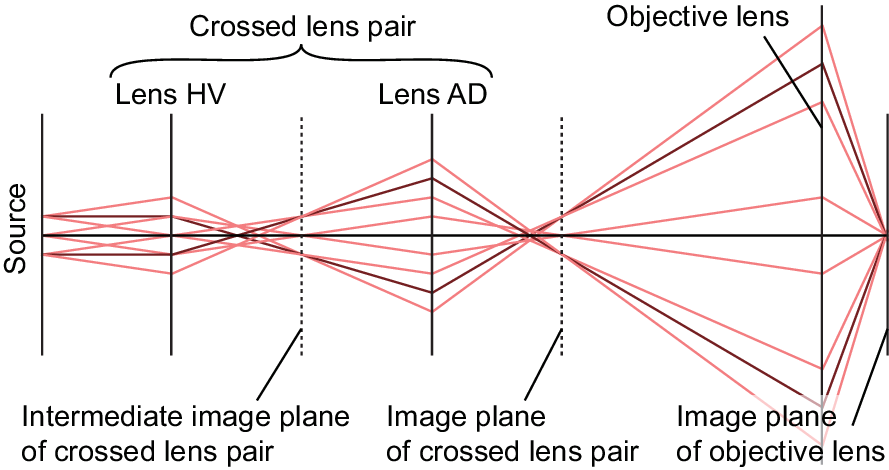}
\caption{Schematic of the electron optical system used to evaluate spherical aberration correction with the crossed lens pair.}
\label{fig:setup}
\end{figure}

As a practical application, we estimate the parameters required to correct third-order spherical aberration in a realistic electron beam probe system using a crossed lens pair.
The optical layout is illustrated in Fig. \ref{fig:setup}.  
An electron beam emitted from an ideal point source with a divergence angle of 0.1 mrad is relayed to the objective lens via the lens HV and the lens AD, each with unity magnification.  
The objective lens, characterized by typical parameters shown in Table \ref{table:objective}, focuses the beam at a 1/100 magnification.  
The divergence angle on the image-side of the objective lens is 10 mrad.  
The ray aberration at the image plane of the objective lens is 1 nm, corresponding to 100 nm at the object plane.  
Accordingly, the required condition for spherical aberration correction is $-1.5\rho = 1.5,C_{\mathrm{s0}}\alpha^3 = -100\,\mathrm{nm}$.
Given $\alpha = 0.1 \mathrm{mrad}$, each crossed lens must provide a circular component of $C_{\mathrm{s0}} = -1.5 \times 10^5$ m.
Assuming the use of $I_{\mathrm{x1}}$ or $I_{\mathrm{xs}}$ for the crossed lenses, the corresponding focal length $f$ and required optical power $P$ are summarized in Table \ref{table:spec}, using Table \ref{table:property}, Eq.(\ref{eq:U}), and Eq.(\ref{eq:I0}).
These results are obtained under the assumptions of a wavelength $\lambda = 1\,\mathrm{\mu m}$, a beam waist $w_0 = 20\,\mathrm{\mu m}$, and an electron energy $T = 1\,\mathrm{keV}$, with the condition $a = 2f$ imposed so that the crossed lens acts as a unity-magnification relay lens.

\begin{table}
    \centering
    \begin{tabular}{ccccc}
    \hline
Magnification & Focal length & 3rd order SA & 5th order SA & 7th order SA \\
    \hline
    \hline
100 & 1 mm & 1 mm & 50 mm & 500 mm \\
    \hline
    \end{tabular}
    \caption{Parameters of the objective lens used to evaluate spherical aberration correction.}
    \label{table:objective}
\end{table}

\begin{table}[ht]
    \centering
    \begin{tabular}{c|c c }
    \hline
    & $f$ & $P$ \\
    \hline
    \hline
    $I_\mathrm{x1}$ &
        7.47 mm &
        33.2 MW \\
    $I_\mathrm{xs}$ &
        555 $\mathrm{\mu}$m &
        30.6 kW \\
    \hline
    \end{tabular}
    \caption{Focal length and input optical power for crossed lenses satisfying $C_{\mathrm{s0}} = -1.5 \times 10^5$ m.}
    \label{table:spec}
\end{table}

\begin{figure}[ht] \centering \includegraphics*[width=0.9\textwidth]{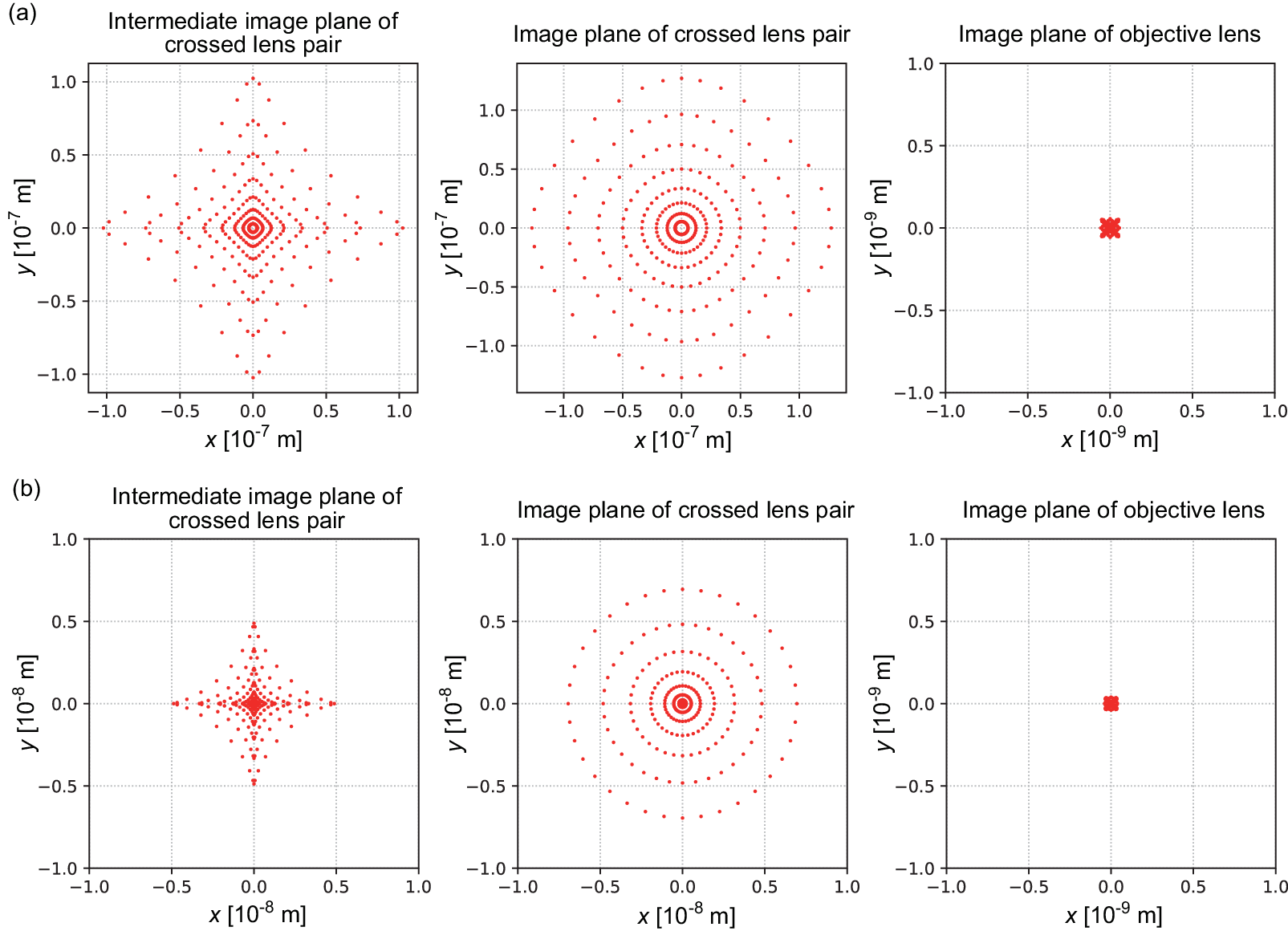} \caption{Spot diagrams of ray aberrations obtained from electron trajectory simulations, using the lens parameters discussed in the main text and the configuration shown in Fig. \ref{fig:setup}. (a) Result for the $I_\mathrm{x1}$ pair. (b) Result for the $I_\mathrm{xs}$ pair.} \label{fig:spots} \end{figure}

To evaluate the performance of the crossed lens pair designed above, we numerically simulated electron trajectories.
The simulation assumes an electron optical approximation in which the electron beam has no energy spread, relativistic effects are neglected (i.e., $\gamma = 1$), the electron energy is conserved, and the velocity is aligned with the propagation axis.
Figure \ref{fig:spots} shows spot diagrams of the ray aberration at the image planes of both crossed lenses and the objective lens.
The incident electrons were assumed to be emitted from the optical axis with divergence angles within a half-angle of 0.1 mrad.
Figure \ref{fig:spots}a corresponds to the $I_\mathrm{x1}$ pair.
We designed the focal length and set it to $0.89f$, slightly shorter than the value in Table \ref{table:spec}, for this simulation.
The resulting minimum spot size at the objective image plane was approximately $5 \times 10^{-10}$ m.
Figure \ref{fig:spots}b shows the result for the $I_\mathrm{xs}$ pair, where a designed focal length of $0.51f$ produced a minimum spot size of about $6 \times 10^{-10}$ m.
In comparison, the uncorrected ray aberration at the objective image plane was $10^{-9}$ m, indicating a spot size reduction by a factor of $\sim20$ in both cases.
Residual aberration is attributed to higher-order spherical aberration in the objective and crossed lenses.

The design formulas for lens power and general spherical aberration coefficients in Table \ref{table:property} are compact and practical, requiring only a minimal set of parameters.
However, since they are derived under the thin-lens approximation, they may deviate from the optimal design values.
This deviation becomes more noticeable when the lens thickness is no longer negligible, such as in short-focal-length lenses.
In the present case, the error in focal length is within 50\%, which is acceptable for practical purposes.
A comparison between crossed lenses constructed from HG beams and those from optical standing waves reveals a three-order-of-magnitude difference in required optical power.
This is primarily due to differences in the steepness of the intensity gradient.
The former, requiring $\sim10$ MW of power, would realistically need to rely on the peak power of ultrashort pulsed lasers.
In contrast, the latter requires only $\sim10$ kW, which may be achieved using a continuous wave laser enhanced by an optical enhancement cavity.
All crossed lenses evaluated in this study are based on the incoherent superposition of two optical beams, which can be implemented using orthogonal polarizations or by axially offsetting the beams along $z$ by more than $2w_0$, without significantly affecting the lens properties.
Of the two crossed lenses formed by coherent superposition of standing waves, $I_\mathrm{dark}$ behaves as a spherical aberration element with no focusing power.
Although it alone cannot generate a negative rotationally symmetric aberration, it could potentially be combined with other crossed lenses to construct an alternative aberration corrector.

\section{Conclusion}
We have derived design formulas for the focal length and general spherical aberration coefficients of crossed ponderomotive lenses.
Using these formulas, we proposed an optical system in which a pair of such lenses can be used to correct third-order spherical aberration in objective lenses for electron microscopy.
Numerical simulations demonstrated that both HG-beam-based and standing-wave-based crossed lens pairs can reduce the electron probe spot size by a factor of approximately 20 for convergence half-angles up to 10 mrad.

Unlike the lens research of the mid-20th century, modern numerical methods enable rapid evaluation of electron optical systems.
Moreover, parameter optimization using machine learning and other algorithmic techniques has reduced the reliance on analytically derived aberration theories.
Nonetheless, compact theoretical frameworks that describe key lens characteristics remain essential as a foundation for system design and understanding.
Active research on aberration theory continues, especially for non-rotationally symmetric systems in applications such as miniaturized imaging optics and extreme ultraviolet focusing optics, where refractive components are not viable \cite{Yuan2009-1,Yuan2009-2,Yuan2009-3,Mori2020,Mori2021,Zhuridov2024-1,Zhuridov2024-2}.
In this context, our analytical expressions provide a useful and accessible tool for the preliminary design and conceptual development of crossed ponderomotive lens systems.

Our results demonstrate that non-rotationally symmetric ponderomotive lenses---based on the perpendicular intersection of optical and electron beams---can be used to design electron optical systems capable of spherical aberration correction.
This represents a significant step toward the practical implementation of ponderomotive lenses and their associated correction techniques in electron microscopy.
Combined with conventional multipole lenses based on electric and magnetic fields, crossed ponderomotive lenses open up new opportunities in electron optical system design.
Such advances may ultimately benefit a broad range of charged-particle technologies, including electron microscopy, photoelectron spectroscopy, and electron beam lithography.

\section*{Acknowledgments}
This work was supported by JST, PRESTO Grant Number JPMJPR2004 and FOREST Grant Number JPMJFR223E.




\end{document}